\documentclass[conference]{IEEEtran}
\IEEEoverridecommandlockouts
\usepackage{cite}
\usepackage{amsmath,amssymb,amsfonts}
\usepackage{subcaption}
\usepackage{graphicx}
\usepackage{caption}
\usepackage{multirow}
\usepackage[Procedure]{algorithm}
\usepackage[noend]{algpseudocode}
\usepackage{graphicx}
\usepackage{textcomp}
\usepackage{xcolor}
\def\BibTeX{{\rm B\kern-.05em{\sc i\kern-.025em b}\kern-.08em
    T\kern-.1667em\lower.7ex\hbox{E}\kern-.125emX}}

\definecolor{orange}{rgb}{1,0.5,0}

\usepackage{ulem}  
\usepackage{url}

\usepackage{booktabs}
\usepackage{makecell}
\usepackage{adjustbox}

\begin{document}

\title{A Test for FLOPs as a Discriminant for Linear Algebra Algorithms\\
\thanks{Financial support from the Deutsche Forschungsgemeinschaft (German Research Foundation) through grants GSC 111 and IRTG 2379 is gratefully acknowledged.}
}

\author{\IEEEauthorblockN{Aravind Sankaran}
\IEEEauthorblockA{\textit{IRTG-MIP}\\
\textit{RWTH Aachen University}\\
Aachen, Germany \\
aravind.sankaran@rwth-aachen.de}
\and
\IEEEauthorblockN{Paolo Bientinesi}
\IEEEauthorblockA{\textit{Department of Computer Science} \\
\textit{Ume\r{a} Universitet}\\
Ume\r{a}, Sweden \\
pauldj@cs.umu.se}


}

\maketitle
\thispagestyle{plain}
\pagestyle{plain}

\begin{abstract}

Linear algebra expressions, which play a central role in countless scientific computations, are often computed via a sequence of calls to existing libraries of building blocks (such as those provided by BLAS and LAPACK). A sequence identifies a computing strategy, i.e., an algorithm, and normally for one linear algebra expression many alternative algorithms exist. Although mathematically equivalent, those algorithms might exhibit significant differences in terms of performance. Several high-level languages and tools for matrix computations such as  Julia, Armadillo, Linnea, etc., make algorithmic choices by minimizing the number of Floating Point Operations (FLOPs). However, there can be several algorithms that share the same (or have nearly identical) number of FLOPs; in many cases, these algorithms exhibit execution times which are statistically equivalent and one could arbitrarily select one of them as the best algorithm. It is however not unlikely to find cases where the execution times are significantly different from one another (despite the FLOP count being almost the same). It is also possible that the algorithm that minimizes FLOPs is not the one that minimizes execution time. In this work, we develop a methodology to test the reliability of FLOPs as discriminant for linear algebra algorithms. Given a set of algorithms (for an instance  of a linear algebra expression) as input, the methodology ranks them into performance classes; i.e., multiple algorithms are allowed to share the same rank. To this end,  we measure the algorithms iteratively until the changes in the ranks converge to a value close to zero. FLOPs are a valid discriminant for an instance if all the algorithms with minimum FLOPs are assigned the best rank; otherwise, the instance is regarded as an anomaly, which can then be used in the investigation of the root cause of performance differences.

\end{abstract}


\section{Introduction}
\label{sec:int}

 One of the major performance bottlenecks for countless computational problems is the evaluation of linear algebra expressions, i.e., expressions involving operations with matrices and/or vectors.  Libraries such as BLAS and LAPACK provide a small set of high performance kernels to compute some standard linear algebra operations~\cite{dongarra1985proposal,demmel1991lapack}. 
However, the mapping of linear  algebra expressions on to an optimized sequence of standard operations is a task far from trivial; the expressions can be computed in many different ways---each corresponding to a specific sequence of library calls---which can significantly differ in performance from one another. Unfortunately, it has been found that the mapping done by most popular high level programming languages such as Matlab, Eigen, TensorFlow, PyTorch, etc., is still suboptimal~\cite{Psarras2022:618,sankaran2022benchmarking}. 

\begin{figure}
	\includegraphics[width=0.5\textwidth]{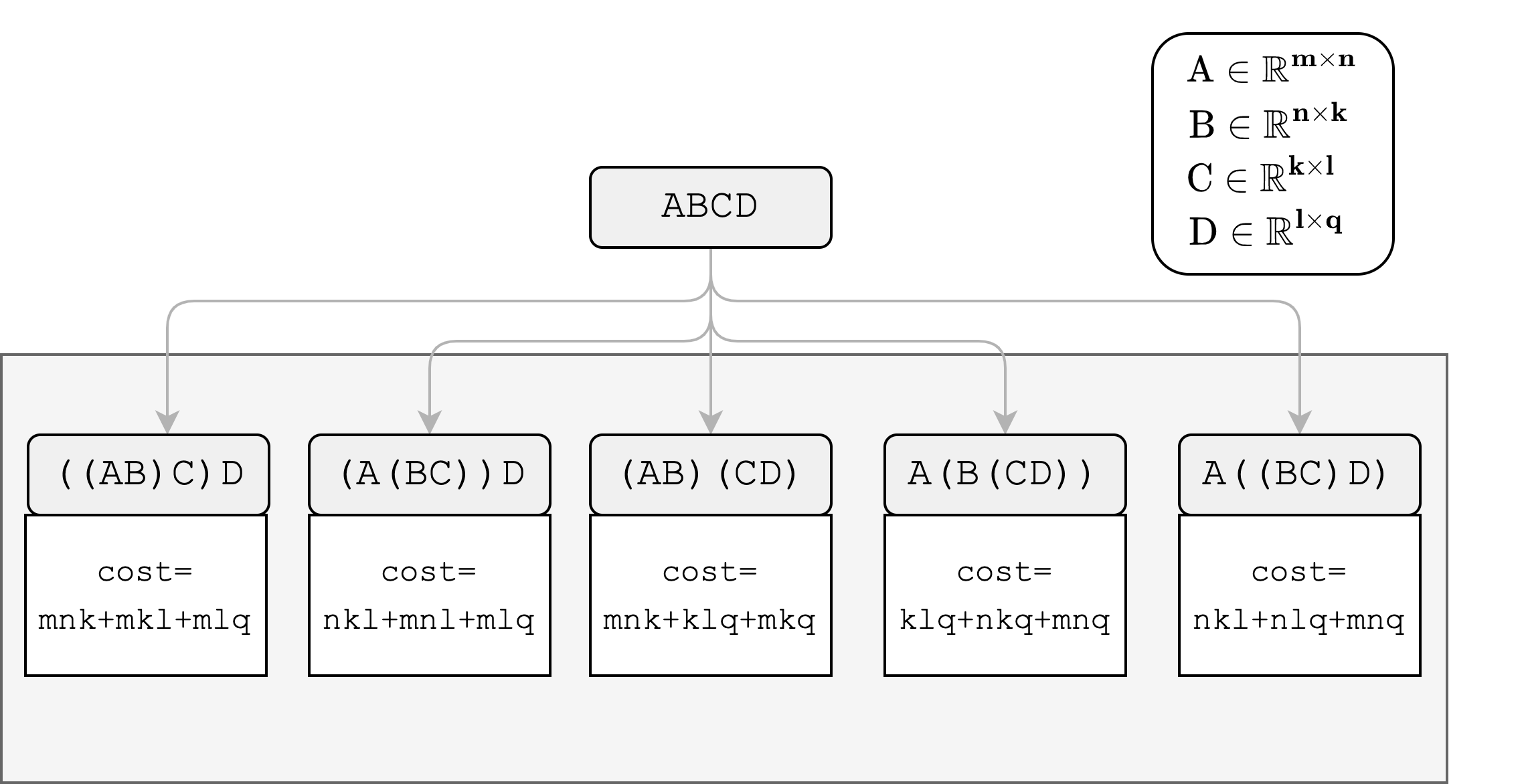}
	\caption{Variants for Matrix chain of length 4. The cost indicates the approximate FLOP count divided by 2.}
	\label{fig:matchain}
\end{figure}

Linear algebra expressions can be manipulated using mathematical properties such as associativity, distributivity, etc., to derive different mathematically equivalent variants (or algorithms). For instance, consider the following expression that evaluates the product of four matrices:
\begin{equation}
\label{eqn:mc}
X = ABCD
\end{equation}
where $A \in \mathbb{R}^{m \times n}$, $B \in \mathbb{R}^{n \times k}$, $C \in \mathbb{R}^{k \times l}$ and $D \in \mathbb{R}^{l \times q}$ are all  dense matrices. An instance of Expression~\ref{eqn:mc} is identified by the tuple $(m,n,k,l,q)$. Because of associativity of the matrix product, Expression~\ref{eqn:mc} can be computed in many different ways, each identified by a specific parenthesization. Although different parenthesizations evaluate to the same mathematical result, they require different number of FLOPs, and might exhibit different performance. For Expression~\ref{eqn:mc}, five possible parenthesizations and their approximate associated costs are shown in Figure~\ref{fig:matchain}.  At least six algorithms can be implemented from the five variants; note that the evaluation of $(AB)(CD)$ can correspond to two different implementations, which differ in the order of instructions, i.e., $AB$ can be computed either before or after $CD$.  A simple strategy to select the fastest algorithm is to select a parenthesization that performs the least floating point operations (FLOPs).  However, it has been observed that the algorithm with the lowest FLOP count is not always the fastest algorithm; such instances are referred to as \textit{anomalies}~\cite{Lopez2022:530}. 

\begin{figure}
	\centering
	\begin{subfigure}[b]{0.5\textwidth}
		\includegraphics[width=1\linewidth]{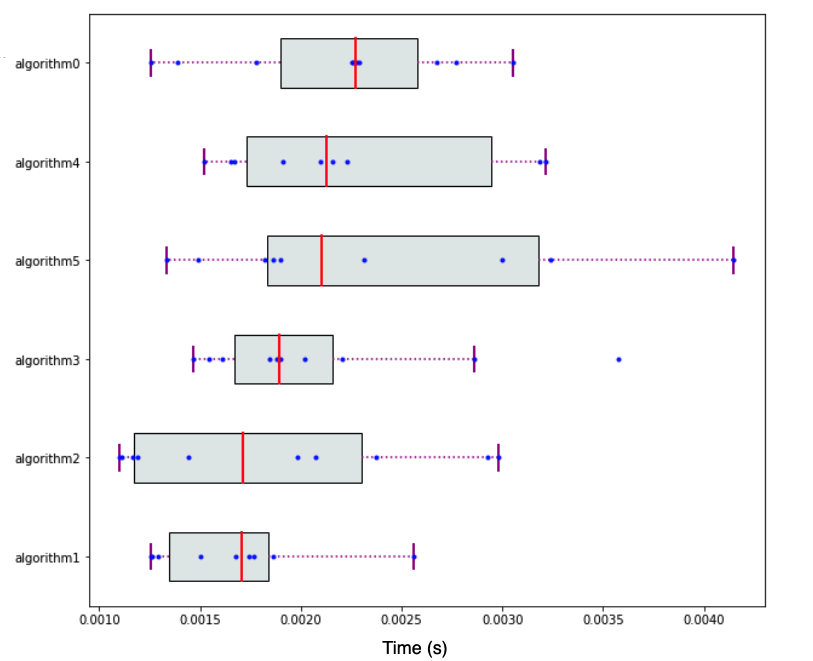}
		\caption{Run 1}
		\label{fig:run1} 
	\end{subfigure}
	\par\bigskip 
	\begin{subfigure}[b]{0.5\textwidth}
		\includegraphics[width=1\linewidth]{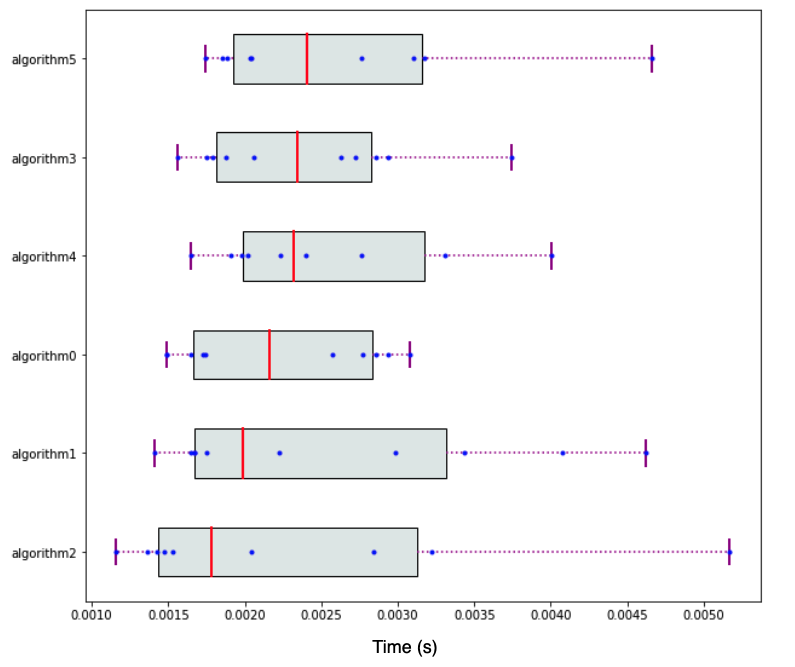}
		\caption{Run 2}
		\label{fig:run2}
	\end{subfigure}
	
	\caption{Two independent runs consisting of 10 measurements of each algorithm for the instance $(331,279,338,854,497)$ of Expression~\ref{eqn:mc}. The algorithms in (a) and (b) are sorted based on increasing median execution times from bottom to top.  }
	\label{fig:rank-medians}
\end{figure}

Consider the following instance of Expression~\ref{eqn:mc}: $(331,279,338,854,497)$, which was observed as an anomaly in~\cite{Lopez2022:530};  there, the algorithms were implemented in C, linked against Intel Math Kernel library\footnote{MKL version 2019.0.5} and measurements were conducted on a Linux-based system using 10 cores of an Intel Xeon processor.
The measurement of each algorithm was repeated 10 times and the median was used to compare algorithms.  Now, in a comparable compute environment, we re-implement the algorithms in Julia\footnote{Julia version 1.3.0}, where the possible influence of the library overheads on execution times can be greater than that of the equivalent C implementations. We link against the same Intel MKL versions and measure the algorithms. The box-plot of the measurements from two independent runs are shown in Figure~\ref{fig:run1} and \ref{fig:run2}. The red line in the box-plot of each algorithm represents the median execution time; the range of the grey box indicates from 25th to 75th quantile, and the length of this box is the Inter-Quartile Region(IQR); the dotted lines are the ``whiskers'' that extend to the smallest and largest observations that are not  outliers according to the 1.5IQR rule~\cite{hodge2004survey}. The difference in FLOP count of an algorithm ($\mathbf{alg}_i$) from the one that computes the least FLOPs is quantified by the Relative FLOPs score (RF$_i$):  
\begin{equation}
\label{eq:rel-flops}
\text{RF}_i =  \frac{F_i - F_{\text{min}}}{F_{\text{min}}}
\end{equation}
where $F_i$ is the FLOPs computed by $\mathbf{alg}_i$ and $F_\text{min}$ is the cost corresponding to the algorithm that computes the least FLOPs. The relative FLOPs scores are shown in Table~\ref{tab:rank-med}.  
\begin{table}[h!]
	\begin{center}
		\renewcommand{\arraystretch}{1.2}
		\begin{adjustbox}{width=1.0\columnwidth,center}
			\begin{tabular}{@{}rr cccccc@{}}
				\toprule
				\textbf{Rank} && \textbf{1} & \textbf{2} & \textbf{3} & \textbf{4} & \textbf{5} & \textbf{6}\\
				\midrule
				Run 1 && \makecell{algorithm1 \\ (0.0)} &  \makecell{algorithm2 \\ (0.04)} &  \makecell{algorithm3 \\ (0.11)} &   \makecell{algorithm5 \\ (0.32)} &  \makecell{algorithm4 \\ (0.27)} &  \makecell{algorithm0 \\ (0.0)}  \\
				Run 2 && \makecell{algorithm2 \\ (0.04)} &  \makecell{algorithm1 \\ (0.0)} &  \makecell{algorithm0 \\ (0.0)} &   \makecell{algorithm4 \\ (0.27)} &  \makecell{algorithm3 \\ (0.11)} &  \makecell{algorithm5 \\ (0.32)}  \\
				\bottomrule
			\end{tabular}
		\end{adjustbox}
		\caption{Algorithms are ranked according to increasing median execution time. The relative FLOP score of every algorithm is indicated in the parenthesis.}
		\label{tab:rank-med}
	\end{center}
\end{table}

It is well known that execution times are influenced by many factors, and that repeated measurements, even with the same input data and compute environment, often result in different execution times~\cite{hoefler2010characterizing, peise2014study}. Therefore, comparing the performance of any two algorithms involves comparing two sets of measurements. In common practice, time measurements are summarized to statistical estimates (such as minimum or median execution time, possibly in combination with standard deviations or quantiles), which are then used to compare and rank algorithms~\cite{Lopez2022:530, hoefler2010characterizing}. However, when the turbo boost settings or inter-kernel cache effects begin to have a significant impact on program execution, the common statistical quantities cannot reliably capture the profile of the time measurements~\cite{hoefler2015scientific}; as a consequence, when time measurements are repeated, the ranking of algorithms would most likely change and this makes the development of reliable performance models for automatic algorithm comparisons difficult.

 Consider again the examples in Figure~\ref{fig:rank-medians}. It can be seen that the ranking of algorithms based on medians are completely different for the two runs. Moreover, the algorithms are not ranked based on the increasing  FLOP counts; in the first run, algorithm0, which is one of the best algorithms in terms of FLOP count (i.e., RF$_1 = 0.0$) is ranked last, and in the second run, algorithm2, which is not among the best algorithms in terms of FLOPs (i.e., RF$_2 \neq 0.0$) is ranked first. The lack of consistency in ranking stems from not considering the possibility that two algorithms can be equivalent when comparing their performances. The box-plots in Figure~\ref{fig:rank-medians} show that the underlying distribution of measurements of the algorithms largely overlap. This indicates that the algorithms with least FLOPs, even though not ranked first,  could have simply been ranked as being equivalent to (as good as) the algorithm chosen as the fastest based on median execution times.
Hence, we elaborate on the definition of anomalies. Let $S_F$ be the set of all algorithms with the least FLOP count. In order to classify an instance as an anomaly, the following conditions are checked one after the other: 
\begin{enumerate}
	\item There should exist an algorithm that is not in $S_F$, but exhibits \textit{noticeably} better performance  than those in $S_F$; that is, we first check if  $S_F$ is a valid representative of the fastest algorithms.
	\item If an instance is not classified as an anomaly  according to (1),  then it is checked if one of the algorithms in $S_F$ exhibit noticeable difference in performance from the rest in $S_F$; this implies, even though   $S_F$ is a valid representative of the fastest algorithms, it is not possible to randomly choose an algorithm from $S_F$ as the best algorithm.
\end{enumerate}

\begin{figure}
	\includegraphics[width=0.5\textwidth]{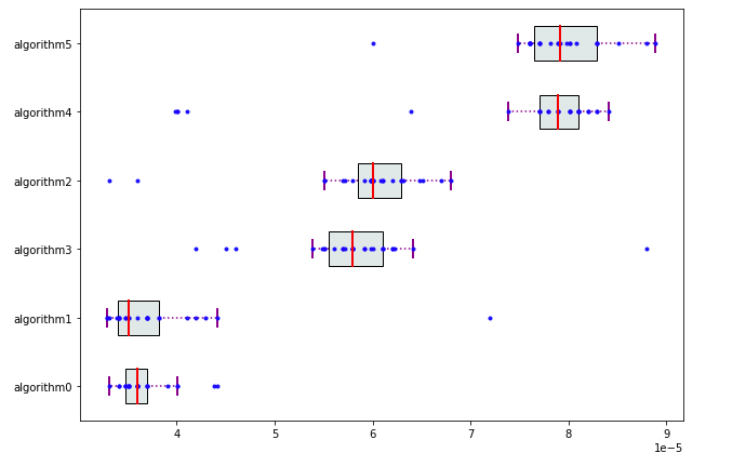}
	\caption{20 measurements of each algorithm for the instance $(75,75,8,75,75)$ of Expression~\ref{eqn:mc}. }
	\label{fig:clusters}
\end{figure}

In order for one algorithm to be faster (or slower) than another, there should be noticeable difference in the distribution of their time measurements; for example, consider another instance of Expression~\ref{eqn:mc}: $(75, 75, 8, 75, 75)$. The measurements of the variant algorithms in Julia are shown in Figure~\ref{fig:clusters}. One could visually infer that algorithms 0 and 1 are equivalent and belong to the same (and best) performance class, while algorithms 2 and 3 have noticeable difference in performance from algorithms 0 and 1, hence belong to a different performance class. The expected ranks for the algorithms are shown in Table~\ref{tab:rank-clusters}. To this end, the comparison of any two algorithms should be able to yield one of the three outcomes: faster, slower, or equivalent. In this paper, we define a three-way comparison function, and develop a methodology that uses this three-way comparison to sort a set of algorithms into performance classes by merging the ranks of algorithms whose distribution of time measurements are significantly overlapping with one another. 
\begin{table}[h!]
	\begin{center}
		\renewcommand{\arraystretch}{1.2}
		\begin{adjustbox}{width=1.0\columnwidth,center}
			\begin{tabular}{@{}ll cccccc@{}}
				\toprule
				\textbf{Algorithm} &&   \makecell{algorithm0 \\ (0.0)} &  \makecell{algorithm1 \\ (0.0)}  &  \makecell{algorithm2 \\ (2.78)}  &  \makecell{algorithm3 \\ (2.78)} &  \makecell{algorithm4 \\ (5.59)}  & \makecell{algorithm5 \\ (5.59)} \\
				\midrule
				Expected rank && 1 & 1&  2& 2&  3& 3\\
				\bottomrule
			\end{tabular}
		\end{adjustbox}
		\caption{Expected ranks for the Algorithms based on the measurements in Figure~\ref{fig:clusters}. The relative FLOP score of every algorithm is indicated in the parenthesis.}
		\label{tab:rank-clusters}
	\end{center}
\end{table}

In practice, linear algebra expressions can have 100s of possible variants. A Statistically sound algorithm comparison requires several repetition of measurements for each variant. As it would be time-consuming to measure all the variants several times, the following approach is employed: 

\begin{enumerate}
	\item After a small warm up to exclude library overheads, all the algorithms are measured exactly once. 
	\item The difference in the execution time of $\mathbf{alg}_i$ from the algorithm with the lowest execution time is quantified by the  Relative Time score (RT$_{i}$):
	\begin{equation}
	\label{eq:rel-time}
	\text{RT}_i =  \frac{T_i - T_{\text{min}}}{T_{\text{min}}}
	\end{equation}
	where $T_i$ is the execution time of $\mathbf{alg}_i$ and $T_\text{min}$ is the minimum observed execution time. 
	\item A set of candidates (S) is created by first shortlisting all the algorithms with minimum FLOP count. Then, to this set, are added all the algorithms that perform more FLOPs, but have relative execution time within a user-specified threshold. 
	\item An initial hypothesis is formed by ranking the candidates in S based on the single-run execution times. 
	\item Each candidate in set S is measured $M$ times (where $M$ is small; e.g., 2 or 3) and ranks of the candidates are updated or merged using the three-way comparison function.  
	\item Step 5  is repeated until the changes in ranks  converge  or the maximum allowed measurements per algorithm  (specified by the user) has been reached. 
\end{enumerate}
If FLOPs are a valid discriminant for a given instance of an expression, then all the algorithms with the least amount of FLOPs would obtain the best rank. Otherwise, the instance would be classified as an anomaly. The identified anomalies can be used to investigate the root cause of performance differences, which would in turn help in the development of meaningful performance models to predict the best algorithm without executing them.

\textit{\textbf{Organization}}: In Sec.~\ref{sec:rel}, we present related works. In
Sec.~\ref{sec:met}, we introduce the methodology to rank algorithms using the three-way comparison. The working of our methodology and the experiments are explained in Sec.~\ref{sec:exp}.  Finally, in Sec.~\ref{sec:con}, we draw conclusions. 
 
\section{Related Works}
\label{sec:rel}

The problem of mapping one target linear algebra expression to a sequence of library calls is known as the Linear Algebra Mapping Problem (LAMP)~\cite{Psarras2022:618}; typical problem instances have many mathematically equivalent solutions, and high-level languages and environments such as Matlab, Julia etc., ideally should select the fastest one. However, it has been shown that most of these languages choose algorithms that are sub-optimal in terms of performance~\cite{Psarras2022:618,sankaran2022benchmarking}. A general approach to identify the fastest algorithm is by ranking the solution algorithms according to their predicted performance. For linear algebra computations, a common performance metric to be used as performance predictor is the FLOP count; however, it has been observed that the number of FLOPs is not always a direct indicator of the fastest code, especially when the computation is bandwidth-bound or executed in parallel~\cite{konstantinidis2015practical, Barthels2021:688}. For selected bandwidth-bound operations, Iakymchuk et al. developed analytical performance models based on memory access patterns~\cite{iakymchuk2011execution, iakymchuk2012modeling}; while those models capture the program execution accurately, their construction requests not only a deep understanding of the processor, but also of the details of the implementation.

There are many examples where an increase in FLOPs count results in a decrease in execution time (anomalies);  ~\cite{bischof1987wy,bischof1994parallel, buttari2008parallel} expose some specific mathematical operations for which the need for complex performance models (that mostly require measuring the execution times) are justified.  However, that does not mean that FLOPs counts are ineffectual for general purpose linear algebra computations that are targeted by the high-level languages. For instance,  in~\cite{sankaran2022benchmarking}, Sankaran et al. expose optimization opportunities in Tensorflow and PyTorch to improve algorithm selection by simply calculating the FLOPs count and applying linear algebra knowledge. In order to justify the need for complex performance models for algorithm selection, it is important to quantify the presence of anomalies. In~\cite{Lopez2022:530}, Lopez et al. estimate the percentage of anomalies for instances of Expression~\ref{eqn:mc} on certain single node multi-threaded setting to be 0.4 percent; in other words, for that case, complex performance models are pointless unless they achieve an accuracy greater than 99.6 percent. It was indicated that the percentage of anomalies increases for more complex expressions. However, in that study, the algorithms were compared using the median execution time from 10 repetition of measurements, and because of this, their comparisons may not be consistent when the experiments are repeated. 

Performance metrics that are a summary of execution times (such as minimum, median etc.) lack in consistency when the measurements of the programs are repeated; this is due to system noise~\cite{hoefler2010characterizing}, cache effects~\cite{peise2014study}, behavior of collective communications~\cite{agarwal2005impact} etc., and it is not realistic to eliminate the performance variations entirely~\cite{alcocer2015tracking}. The distribution of execution times obtained by repeated measurements of a program is known to lack in textbook statistical behaviours and hence specialized methods to quantify performance have been developed~\cite{chen2014statistical, chen2016robust, hoefler2010loggopsim, bohme2016identifying}. However, approximating statistical distributions require executing the algorithms several times. In this work, we develop a strategy to minimize the number of measurements.

The performance of an algorithm can be predicted using regression or machine learning based methods; this requires careful formulation of an underlying problem. A wide body of significant work has been done in this direction for more than a decade~\cite{peise2014performance, barnes2008regression, barve2019fecbench}. Peise et al in~\cite{peise2014performance} create a prediction model for individual BLAS calls and estimates the execution time for an algorithm by composing the predictions from several BLAS calls. In~\cite{barve2019fecbench}, Barve et al predict performance to optimize resource allocation in multi-tenant systems. Barnes et al in~\cite{barnes2008regression} predict scalability of parallel algorithms. In those approaches, the performances are quantified in an absolute term. Instead, in this work, we quantify performance of algorithms relatively to one another using pairwise comparisons. In~\cite{sankaran2021performance}, Sankaran et al. discuss an application of algorithm ranking via relative performance in an edge computing environment to reduce energy consumption in devices. They compare algorithms by randomly bootstrapping measurements. Instead, we present an approach that compares algorithms by comparing the quantiles from the measurements.

\section{Methodology}
\label{sec:met}

Let  $\mathcal{A}$  be  a set of mathematically equivalent  algorithms. The algorithms  $\mathbf{alg}_1, \dots, \mathbf{alg}_p \in \mathcal{A}$ are ordered according to decreasing performance based on some initial hypothesis.  Let  $\mathbf{h_0}: \langle \mathbf{alg}_{i(1)}, \dots, \mathbf{alg}_{i(j)}, \dots \mathbf{alg}_{i(p)} \rangle$ be an initial ordering. Here, $i(j)$ is the index of the algorithm at position  $j$ in $\mathbf{h_0}$.  For instance, consider the equivalent algorithms in Figure~\ref{fig:clusters}: $\{ \mathbf{alg}_0, \mathbf{alg}_1,\mathbf{alg}_2, \mathbf{alg}_3, \mathbf{alg}_4,\mathbf{alg}_5  \}$. If the initial hypothesis is formed based on the increasing \textit{minimum} execution times observed for each algorithm, then the initial ordering would be  $\mathbf{h_0}:  \langle \mathbf{alg}_2, \mathbf{alg}_1, \mathbf{alg}_0, \mathbf{alg}_4, \mathbf{alg}_3, \mathbf{alg}_5 \rangle $. Here, the index of the algorithm in the first  position is $i(1)=2$, the second position is $i(2)=1$, and so on. The execution time of each algorithm in $\mathbf{h_0}$ is measured $N$ times, and based on the additional empirical evidence, the algorithms are re-ordered to produce a sequence consisting of tuples $\mathbf{s}: \langle (\mathbf{alg}_{s(1)},\textbf{rank}_1), \dots, (\mathbf{alg}_{s(p)},\textbf{rank}_p) \rangle$,  where $\textbf{rank}_j$ is the rank of the algorithm at position $j$ in $\mathbf{s}$ and $\textbf{rank}_j \in \{1, \dots k\}$ with $k \le  p$ (i.e., several algorithms can share the same rank).  Here, $s(j)$ is the index of the algorithm at $j^{th}$ position in $\mathbf{s}$. For instance, for the experiment in Figure~\ref{fig:clusters}, the sorted sequence would be  $\langle (\mathbf{alg}_{1},1), (\mathbf{alg}_{0},1), (\mathbf{alg}_{2},2), (\mathbf{alg}_{3},2), (\mathbf{alg}_{4},3), (\mathbf{alg}_{5},3) \rangle$. Algorithms that evaluate to be equivalent to one another are assigned the same rank. To this end, we first  define the procedure to compare two algorithms that takes into account the equivalence of the algorithms. Then, we sort the algorithms using the three-way comparison function to update and merge ranks.

\textbf{Algorithm comparison (Procedure~\ref{alg:compare}):} Procedure~\ref{alg:compare} takes in as input any two sets of $N$ measurements $\mathbf{t}_i, \mathbf{t_j} \in \mathbb{R}^N$ from algorithms $\mathbf{alg}_i, \mathbf{alg}_j$ respectively, and a specific quantile  range  $(q_{\text{lower}}, q_{\text{upper}})$. The procedure compares the quantiles of the two algorithms. If the $ q_{\text{upper}}$ of $\mathbf{alg}_i$  is less than the $ q_{\text{lower}}$ of $\mathbf{alg}_j$, then  $\mathbf{alg}_i$ is ``better'' than ($<$) $\mathbf{alg}_j$. Otherwise, if  $ q_{\text{upper}}$ of $\mathbf{alg}_j$  is less than the $ q_{\text{lower}}$ of $\mathbf{alg}_i$, then  $\mathbf{alg}_i$ is ``worse'' than ($>$) $\mathbf{alg}_j$. Otherwise, both the  algorithms are ``equivalent''($\sim$).

\begin{algorithm}
	\caption{CompareAlgs $(\mathbf{alg}_i, \mathbf{alg}_j, q_{\text{lower}}, q_{\text{upper}})$ }
	\label{alg:compare}
	\hspace*{\algorithmicindent} \textbf{Inp: } $ \mathbf{alg}_i, \mathbf{alg}_j \in \mathcal{A}$ \\ 
	\hspace*{\algorithmicindent} \hspace*{\algorithmicindent}  $  \quad q_{\text{lower}},  q_{\text{upper}} \in (0,100) \quad q_{\text{upper}} > q_{\text{lower}}  $\\ 
	\hspace*{\algorithmicindent} \textbf{Out:} $  \{``\mathbf{alg}_i < \mathbf{alg}_j", ``\mathbf{alg}_i>\mathbf{alg}_j", ``\mathbf{alg}_i\sim \mathbf{alg}_j" \}$
	\begin{algorithmic}[1] 
		\State $\mathbf{t_i} = get\_measurements(\mathbf{alg}_i)$ \Comment{$\mathbf{t_i} \in \mathbb{R}^{N}$}
		\State $\mathbf{t_j} = get\_measurements(\mathbf{alg}_j)$ \Comment{$\mathbf{t_j} \in \mathbb{R}^{N}$} \\
		\State $\mathbf{t_i}^{low} \leftarrow $ Value of the $q_{\text{lower}}$ quantile in $\mathbf{t_i}$ \Comment{$ \mathbf{t_i}^{low} \in \mathbb{R}$}
		\State $ \mathbf{t_i}^{up} \leftarrow $ Value of the $q_{\text{upper}}$ quantile in $\mathbf{t_i}$ \Comment{$ \mathbf{t_i}^{up} \in \mathbb{R}$}\\
		\State $\mathbf{t_j}^{low} \leftarrow $ Value of the $q_{\text{lower}}$ quantile in $\mathbf{t_j}$ \Comment{$ \mathbf{t_j}^{low} \in \mathbb{R}$}
		\State $ \mathbf{t_j}^{up} \leftarrow $ Value of the $q_{\text{upper}}$ quantile in $\mathbf{t_j}$  \Comment{$ \mathbf{t_j}^{up} \in \mathbb{R}$}\\
		\If{$\mathbf{t_i}^{up}  < \mathbf{t_j}^{low} $}
		\State return ``$\mathbf{alg}_i < \mathbf{alg}_j$"
		\ElsIf{$\mathbf{t_j}^{up}  < \mathbf{t_i}^{low} $}
		\State return ``$\mathbf{alg}_i>\mathbf{alg}_j$" 
		\Else
		\State return ``$\mathbf{alg}_i\sim \mathbf{alg}_j $"
		\EndIf
	\end{algorithmic}
\end{algorithm}

\begin{figure}
	\includegraphics[width=0.5\textwidth]{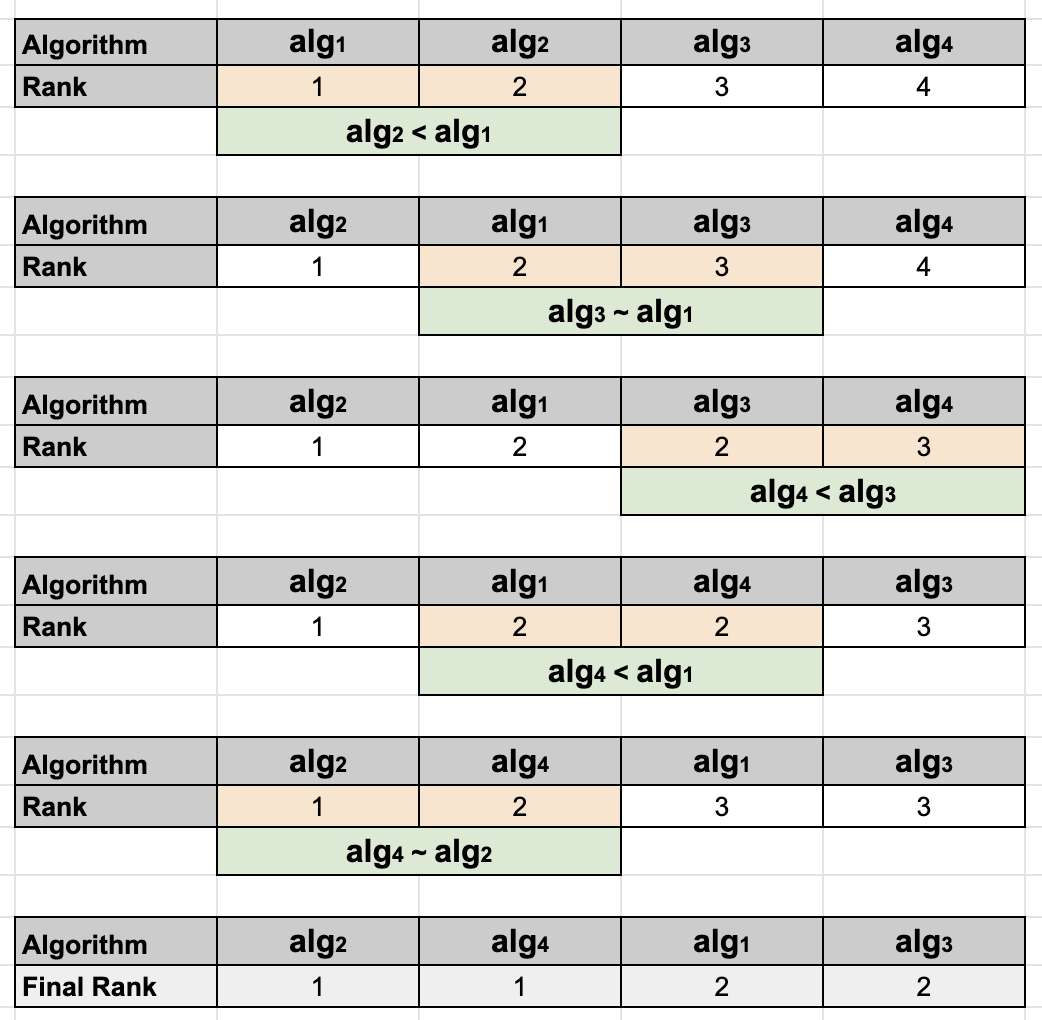}
	\caption{Bubble Sort with the three-way comparison function.}
	\label{fig:sort}
\end{figure}

\textbf{Sorting procedure (Procedure~\ref{alg:sort}):} The inputs to  Procedure~\ref{alg:sort} are the initial sequence $\mathbf{h}_0$ and a quantile  range $(q_{\text{lower}}, q_{\text{upper}})$. The output is a sorted sequence set $\mathbf{s}$. To this end, the bubble-sort procedure~\cite{astrachan2003bubble} is adapted to work with the three-way comparison function. Starting from  the left most element in the initial sequence,  the procedure compares adjacent algorithms and swaps their positions if an algorithm occurring later in the sequence is better (according to Procedure~\ref{alg:compare})  than the previous algorithm, and then ranks are updated. When the comparison of two algorithms results to be equivalent as each other, both are assigned with the same rank, but their positions are not swapped. In order to illustrate the rank update rules in detail, we consider the illustration in Figure~\ref{fig:sort}, which shows the intermediate steps while sorting an initial sequence  $  \mathbf{h_0}: \langle \mathbf{alg}_1, \mathbf{alg}_2, \mathbf{alg}_3, \mathbf{alg}_4 \rangle $. All possible update rules that one might encounter appear in one of the intermediate steps of this example.
\begin{enumerate}
	\item 
	\textit{\textbf{Both positions and ranks are swapped} :}
	In the first pass of bubble sort, pair-wise comparison of adjacent algorithms are done starting from the first
	element in the sequence. Currently, the sequence is $\langle \mathbf{alg}_1, \mathbf{alg}_2, \mathbf{alg}_3, \mathbf{alg}_4 \rangle$.
	As a first step, algorithms $\mathbf{alg}_1$ and $\mathbf{alg}_2$ are compared, and 
	$\mathbf{alg}_2$ ends up being faster. As the slower algorithm should be shifted towards the end of the sequence, $\mathbf{alg}_1$ and $\mathbf{alg}_2$ swap positions  (line \ref{lst:swap} in
	Procedure~\ref{alg:sort}). Since all the algorithms still have unique ranks, $\mathbf{alg}_2$ and
	$\mathbf{alg}_1$ also exchange their ranks, and no special rules for updating ranks are applied.
	So, $\mathbf{alg}_2$ and $\mathbf{alg}_1$ receive rank 1 and 2, respectively.
	
	\item \textit{\textbf{Positions are not swapped but the ranks are merged}:} Next, algorithm $\mathbf{alg}_1$ is compared with its successor $\mathbf{alg}_3$;
	since they are just as good as one another, 
	no swap takes place. Now, the rank of $\mathbf{alg}_3$ should also indicate that it is as good as $\mathbf{alg}_1$;
	so $\mathbf{alg}_3$ is given the same rank as $\mathbf{alg}_1$ 
	and the rank of $\mathbf{alg}_4$ is corrected by
	decrementing by 1. (line \ref{lst:ag1}-\ref{lst:ag2} in Procedure \ref{alg:sort}). Hence $\mathbf{alg}_1$ and $\mathbf{alg}_3$ have rank 2, and $\mathbf{alg}_4$ is corrected to rank 3.
	
	\item
	\textit{\textbf{Both positions and ranks are swapped}:} (\textit{This is the same rule applied in Step 1}).
	In the last comparison of the first sweep of bubble sort, Algorithm $\mathbf{alg}_4$ results to be faster than $\mathbf{alg}_3$, so their positions and ranks are
	swapped. 
	This completes the first pass of bubble-sort. At this point, the sequence is $\langle \mathbf{alg}_2, \mathbf{alg}_1, \mathbf{alg}_4, \mathbf{alg}_3 \rangle$.
	
	\item
	\textit{\textbf{Swapping positions with algorithms having same rank}:} In the second pass of bubble sort, the pair-wise comparison of adjacent algorithms, except the ``right-most" algorithm in sequence, is evaluated (note that the right-most algorithm can still have its rank updated depending upon the results of comparisons of algorithms occurring earlier in the sequence). The first two algorithms $\mathbf{alg}_2$ and
	$\mathbf{alg}_1$ were already compared in Step 1. So now, the next comparison is $\mathbf{alg}_1$ vs.~$\mathbf{alg}_4$.
	Algorithm $\mathbf{alg}_4$ results to be faster than $\mathbf{alg}_1$ although they were assigned the same rank. Therefore, their positions are swapped as usual, but the
	rank of $\mathbf{alg}_4$ remains the same and only the rank of $\mathbf{alg}_1$ is incremented by 1.
	(line \ref{lst:h1}-\ref{lst:h2} in Procedure \ref{alg:sort}).
	  This completes the second pass of bubble sort and the two slowest algorithms have been pushed to the right.
	
	\item
	\textit{\textbf{Positions are not swapped but the ranks are merged:}} (\textit{This is the same rule applied in Step 2}). In the third and final pass,
	we again start from the first element on the left of the sequence and continue the pair-wise comparisons until the third last
	element. This leaves only one comparison to be done, $\mathbf{alg}_4$ vs.~$\mathbf{alg}_2$.
	Algorithm $\mathbf{alg}_4$ is evaluated to be as good as $\mathbf{alg}_2$, so both are given the same rank and the positions are not swapped. The ranks of algorithms occurring later than $\mathbf{alg}_4$ in the sequence are decremented by 1. Thus,
	the final sequence is $\langle \mathbf{alg}_2, \mathbf{alg}_4, \mathbf{alg}_1, \mathbf{alg}_3 \rangle$. Algorithms $\mathbf{alg}_2$ and $\mathbf{alg}_4$ obtain rank 1, and $\mathbf{alg}_1$ and $\mathbf{alg}_3$ obtain rank 2.
\end{enumerate}
\begin{algorithm}
	\caption{SortAlgs $(\mathbf{h_0}, q_{\text{lower}}, q_{\text{upper}})$ }
	\label{alg:sort}
	\hspace*{\algorithmicindent} \textbf{Input: } $ \mathbf{h_0} : \langle \mathbf{alg}_{i(1)},\dots,\mathbf{alg}_{i(p)} \rangle $ \\
	\hspace*{\algorithmicindent} \hspace*{\algorithmicindent}  $  \quad q_{\text{lower}},  q_{\text{upper}} \in (0,100) \quad q_{\text{upper}} > q_{\text{lower}}  $\\ 
	\hspace*{\algorithmicindent} \textbf{Output: } $ \mathbf{s}: \langle (\mathbf{alg}_{s(1)},r_1), \dots, (\mathbf{alg}_{s(p)},r_p) \rangle $
	\begin{algorithmic}[1] 
		\For{j = 1, $\dots$, p}
		\State  Initialize $r_j \leftarrow j$ \Comment{Initialize Alg rank}
		\State  Initialize $s(j) \leftarrow i(j)$ \Comment{Initialize Alg order}
		\EndFor\\
		\For{k = 1, $\dots$, p}
		\For{j = 0, $\dots$, p-k-1}
		\State ret = CompareAlgs($\mathbf{alg}_{s(j)}, \mathbf{alg}_{s(j+1)},  q_{\text{lower}}, q_{\text{upper}}$)
		\If{$\mathbf{alg}_{s(j+1)}$ is faster than $\mathbf{alg}_{s(j)}$}
		\State Swap indices $s(j)$ and $s(j+1)$ \label{lst:swap}
		\If{$r_{j+1} = r_j$} \label{lst:h1}
		\State Increment rank $r_{j+1}$ by 1 \label{lst:h2}
		\EndIf
		\ElsIf{$\mathbf{alg}_{s(j+1)}$ is as good as $\mathbf{alg}_{s(j)}$} \label{lst:ag1}
		\If{$r_{j+1} \ne r_j$}
		\State Decrement ranks $r_{j+1}, \dots, r_p$ by 1 \label{lst:ag2}
		\EndIf
		\ElsIf{$\mathbf{alg}_{s(j)}$ is faster than $\mathbf{alg}_{s(j+1)}$}
		\State Leave the ranks as they are
		\EndIf		
		\EndFor
		\EndFor
		\State $\mathbf{s} \leftarrow \langle (\mathbf{alg}_{s(1)},r_1), \dots, (\mathbf{alg}_{s(p)},r_p) \rangle$
		\State return $\mathbf{s}$
	\end{algorithmic}
\end{algorithm}
\bigskip

\textbf{Mean rank calculation (Procedure 3):} The results of the sorting procedure depend on the chosen quantile ranges $(q_{\text{lower}}, q_{\text{upper}})$. For the example in Figure~\ref{fig:clusters}, the estimated ranks for different  quantile ranges are shown in Table~\ref{tab:q-ranks}.  For large quantile ranges that cover the tail ends of the time distribution, such as $(q_5, q_{95})$, the algorithms result to be equivalent to one another more often than the small ranges. For instance, in our example, all the algorithms are estimated to be equivalent (i.e., rank 1) for the quantile  range $(q_5, q_{95})$. As the quantile ranges become smaller and smaller, the tails of the distributions  are curtailed, and the overlaps estimated among the algorithms become lesser and lesser. Thus, for $(q_{25}, q_{75})$, $\mathbf{alg_0}$, $\mathbf{alg_1}$ obtain rank 1, $\mathbf{alg_2}$, $\mathbf{alg_3}$ obtain rank 2, and $\mathbf{alg_4}$, $\mathbf{alg_5}$ obtain rank 3; these are the ranks one might expect according to the visual inference of the box-plots in Figure~\ref{fig:clusters}.   For a smaller range, $(q_{35}, q_{65})$, $\mathbf{alg_2}$ and $\mathbf{alg_3}$ obtain different ranks as $\mathbf{alg_2}$ is slightly shifted towards the right of $\mathbf{alg}_3$ despite  significant overlap. Therefore, ranks from isolated quantile ranges does not accurately quantify the underlying performance characteristics of the algorithms. 
\begin{table}[h!]
	\begin{center}
		\renewcommand{\arraystretch}{1.2}
		\begin{tabular}{@{}r ccc ccc@{}}
			\toprule
			& \textbf{alg1} & \textbf{alg0}  & \textbf{alg3}  & \textbf{alg2} &\textbf{alg4}  & \textbf{alg5}  \\
			\midrule
			{$(q_5, q_{95})$} & 1  & 1  & 1  & 1  & 1  & 1  \\
			{$(q_{10}, q_{90})$} & 1  & 1   & 2  & 2  & 2   & 2  \\
			{$(q_{15}, q_{85})$} & 1  & 1  & 2  & 2  & 2  & 2  \\
			{$(q_{20}, q_{80})$} & 1  & 1  & 2  & 2  & 3  & 3  \\
			{$\mathbf{(q_{25}, q_{75})}$} & \textbf{1}  &\textbf{1}  & \textbf{2}  &\textbf{2}  & \textbf{3}  & \textbf{3}  \\
			{$(q_{30}, q_{70})$} & 1  & 1  & 2  & 2  & 3  & 3  \\
			{$(q_{35}, q_{65})$} & 1  & 1  & 2  & 3  & 4  & 4  \\
			{\textbf{Mean rank}} & \textbf{1.0}   & \textbf{1.0}  & \textbf{1.86 } & \textbf{2.0}  & \textbf{2.57 } & \textbf{2.57 } \\
			\bottomrule
		\end{tabular}
		\caption{Ranks calculated for the data in Figure~\ref{fig:clusters} on different quantile ranges. The mean ranks of the algorithms  across the quantile ranges are also shown. }
		\label{tab:q-ranks}
	\end{center}
\end{table}

In order to estimate a reliable metric, we repeat Procedure~\ref{alg:sort}, and compute ranks with different quantile ranges, and compute the mean rank ($\mathbf{mr}$) for each algorithm. The ranks from a specific quantile range can be compared with the mean ranks to get a better understanding of the performance. We choose $(q_{25}, q_{75})$ as default, as this range is considered as a default for statistical outlier detections~\cite{hodge2004survey}. For $(q_{25}, q_{75})$, both $\mathbf{alg_2}$ and $\mathbf{alg_3}$ obtain the same rank; however, as the mean rank for $\mathbf{alg_2}$ is slightly greater than $\mathbf{alg_3}$, this indicates that according to the available empirical data,   $\mathbf{alg_3}$ is better than  $\mathbf{alg_2}$.

\begin{algorithm}
	\caption{ MeanRanks$(\mathbf{h_0}, \mathbf{q})$ }
	\label{alg:meanrank}
	\hspace*{\algorithmicindent} \textbf{Input: } $ \mathbf{h_0} : \langle \mathbf{alg}_{i(1)},\dots,\mathbf{alg}_{i(p)} \rangle$ \\
	\hspace*{\algorithmicindent} \hspace*{\algorithmicindent}  $  \mathbf{q} : [ (q1_{\text{low}}, q1_{\text{up}}), \dots ,(qH_{\text{low}}, qH_{\text{up}})] $\\ 
	\hspace*{\algorithmicindent} \textbf{Output: } $ \mathbf{s}_{[25,75]}, [(\mathbf{alg}_1, \mathbf{mr}_1), \dots, (\mathbf{alg}_p, \mathbf{mr}_p)]$
	\begin{algorithmic}[1] 
		\State  Initialize $\mathbf{mr}_i \leftarrow 0$ \Comment{$i \in \{1, \dots,  p\}$}
		\For{$q_{\text{low}}$, $q_{\text{up}}$ in $\mathbf{q}$}
		\State $\mathbf{s}_{[low, up]} \leftarrow$ SortAlgs$(\mathbf{h_0}, q_{\text{low}}, q_{\text{up}})$ 
		\EndFor
		\State $\mathbf{mr}_i \leftarrow$ Mean rank  of  $\mathbf{alg}_i$ over all the quantiles  in $\mathbf{s}$.
		\State return $\mathbf{s}_{[25,75]}, [(\mathbf{alg}_1, \mathbf{mr}_1), \dots, (\mathbf{alg}_p, \mathbf{mr}_p)]$
	\end{algorithmic}
\end{algorithm}

\textbf{Convergence (Procedure 4): } The calculation of ranks requires measurements of execution times for each algorithm.  Starting with an empty measurement set (i.e., $N=0$), $M$ measurements (typically only a few;  e.g., 2 or 3) of each algorithm are iteratively added and the mean ranks are computed using Procedure~\ref{alg:meanrank}. The iteration stops as the mean ranks converge. We estimate the convergence of a rank as follows:
Let $\mathbf{x}$ be an ordered list of mean ranks. Then the changes in mean rank between adjacent  algorithms in the list ($\mathbf{dx}$) is computed as:
\begin{equation*}
\mathbf{dx} = \text{convolution}(\mathbf{x}, [1,-1], step=1)
\end{equation*}
where [1,-1] is the convolution filter. In simple words, $\mathbf{dx}$ is the difference in the mean ranks between the adjacent algorithms in $\mathbf{x}$. 
Let $\mathbf{dx}$ and $\mathbf{dy}$ be the changes in mean ranks over  the lists from iteration $j$ and $j-1$ respectively. Then, the stopping criterion for the iteration is 
\begin{equation*}
\frac{\lVert \mathbf{dx} - \mathbf{dy} \rVert_{L2}}{p} < eps
\end{equation*}
 where $\lVert . \rVert_{L2}$ is the L2 norm and $p$ is the number of algorithms being compared. The iterations stop when the stopping criterion becomes less than $eps$ or  if the number of measurements per algorithm reaches a user specified maximum value ($max$). 
 
 For illustration, follow the example in Table~\ref{tab:q-ranks}. The list of mean ranks is $\mathbf{x} = [1,1,1.86,2.0,2.57,2.57]$, and the changes in mean ranks is $\mathbf{dx} = [0,0.86,0.14,0.57,0]$. Now, consider the ranks at $(q_{35}, q_{65})$, the ranks of the algorithms are $\langle (\mathbf{alg}_{1},1), (\mathbf{alg}_{0},1), (\mathbf{alg}_{3},2), (\mathbf{alg}_{2},3), (\mathbf{alg}_{4},4), (\mathbf{alg}_{5},4) \rangle$. Let us assume that in the following iteration, $\mathbf{alg_2}$ and $\mathbf{alg_3}$ obtain the same rank; i.e., $\langle (\mathbf{alg}_{1},1), (\mathbf{alg}_{0},1), (\mathbf{alg}_{3},2), (\mathbf{alg}_{2},2), (\mathbf{alg}_{4},3), (\mathbf{alg}_{5},3) \rangle$, then the mean rank for next iteration would be $\mathbf{y} = [1,1,1.86,1.86,2.43,2.43]$, and the changes in mean rank would be $\mathbf{dy} = [0,0.86,0,0.57,0]$. Then $ \frac{\lVert \mathbf{dy} - \mathbf{dx} \rVert_{L2}}{5} = 0.028$. 
 
 
 \begin{algorithm}
 	\caption{ MeasureAndRank$(\mathbf{h_0}, M, eps, max)$ }
 	\label{alg:measure}
 	\hspace*{\algorithmicindent} \textbf{Input: } $ \mathbf{h_0} :\langle \mathbf{alg}_{i(1)},\dots,\mathbf{alg}_{i(p)} \rangle$ \\
 	\hspace*{\algorithmicindent} \hspace*{\algorithmicindent}  $  M, eps, max \in \mathbb{R} $\\ 
 	\hspace*{\algorithmicindent} \textbf{Output: } $\mathbf{s}_{[25,75]}, \mathbf{mr}'$
 	\begin{algorithmic}[1] 
 		\State Set norm $\gg eps$ \Comment{norm $\in \mathbb{R}$}
 		\State $\mathbf{q} \leftarrow$ Define a set of quantile ranges.
 		\State $N = 0$
 		\State Initialize $\mathbf{dy}_j \leftarrow 1 $ \Comment{$j \in \{1,\dots, p-1\}$}
 		\While{norm $\ge eps$ \textbf{and}  $N < max$}
		\State Measure every algorithm $M$ times.
		\State  $N = N+M$
		\State  $\mathbf{s}_{[25,75]}, \mathbf{mr}' \leftarrow $MeanRanks($\mathbf{h_0}, \mathbf{q}$) 
		\State \Comment{On N Measurements}
		\State $\mathbf{x}_i \leftarrow $  Mean rank of $\mathbf{alg}_i$ from $\mathbf{mr'}$ \Comment{$ i \in  \{1,\dots, p\}$}
		\State $\mathbf{dx} \leftarrow \text{convolution}(\mathbf{x}, [1,-1], step=1)$ \Comment{$\mathbf{x} \in \mathbb{R}^{p}$} 
		\State norm  $ \leftarrow \frac{\lVert \mathbf{dx} - \mathbf{dy} \rVert_{L2}}{p}$ \Comment{$\mathbf{dx} \in \mathbb{R}^{p-1} $}
		\State $\mathbf{dy} \leftarrow \mathbf{dx}$
		\State  $\mathbf{h_0} \leftarrow   \langle \mathbf{alg}_{s(1)},\dots,\mathbf{alg}_{s(p)} \rangle $ \Comment{Ordering from $\mathbf{s}_{[25,75]}$}
		\EndWhile
 		\State return $\mathbf{s}_{[25,75]}, \mathbf{mr}'$
 	\end{algorithmic}
 \end{algorithm}

%
%

\section{Interpretation}
\label{sec:exp}

In this section, we explain the working of our methodology. For our experiments, we use the Linnea framework~\cite{Barthels2021:688}, which accepts linear algebra expressions as input, and generates a  family of  mathematically equivalent algorithms (in the form of sequential Julia code) consisting of (mostly, but not limited to) sequences of BLAS and  LAPACK calls. The experiments are run on a Linux based Intel Xeon machine with turbo-boost enabled and number of threads set to 10. 

\begin{figure}
	\centering
	
	\begin{subfigure}[b]{0.5\textwidth}
		\includegraphics[width=1\linewidth]{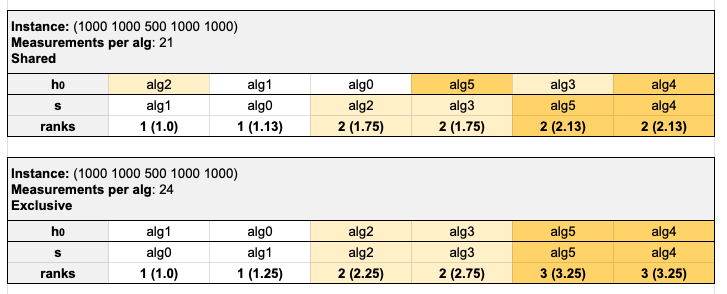}
		\caption{Instance A}
		\label{fig:exp2b}
	\end{subfigure}
	
	\begin{subfigure}[b]{0.5\textwidth}
		\includegraphics[width=1\linewidth]{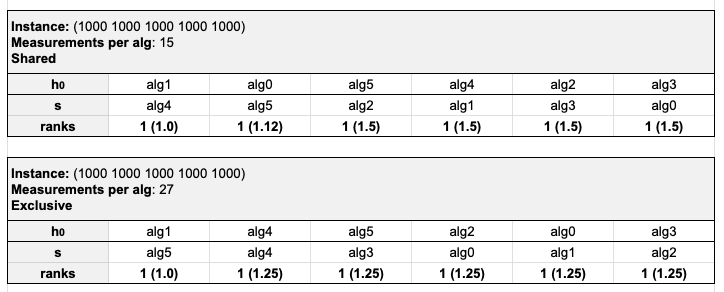}
		\caption{Instance B}
		\label{fig:exp2c}
	\end{subfigure}
	
	\caption{$\mathbf{h_0}$ is the Initial hypothesis. $\mathbf{s}$ is the updated sequence. The ranks at $(q_{25}, q_{75})$ and the mean ranks (in brackets) are shown.}
	\label{fig:exp2}
\end{figure}

\begin{figure}
	\centering
	\begin{subfigure}[b]{0.5\textwidth}
		\includegraphics[width=1\linewidth]{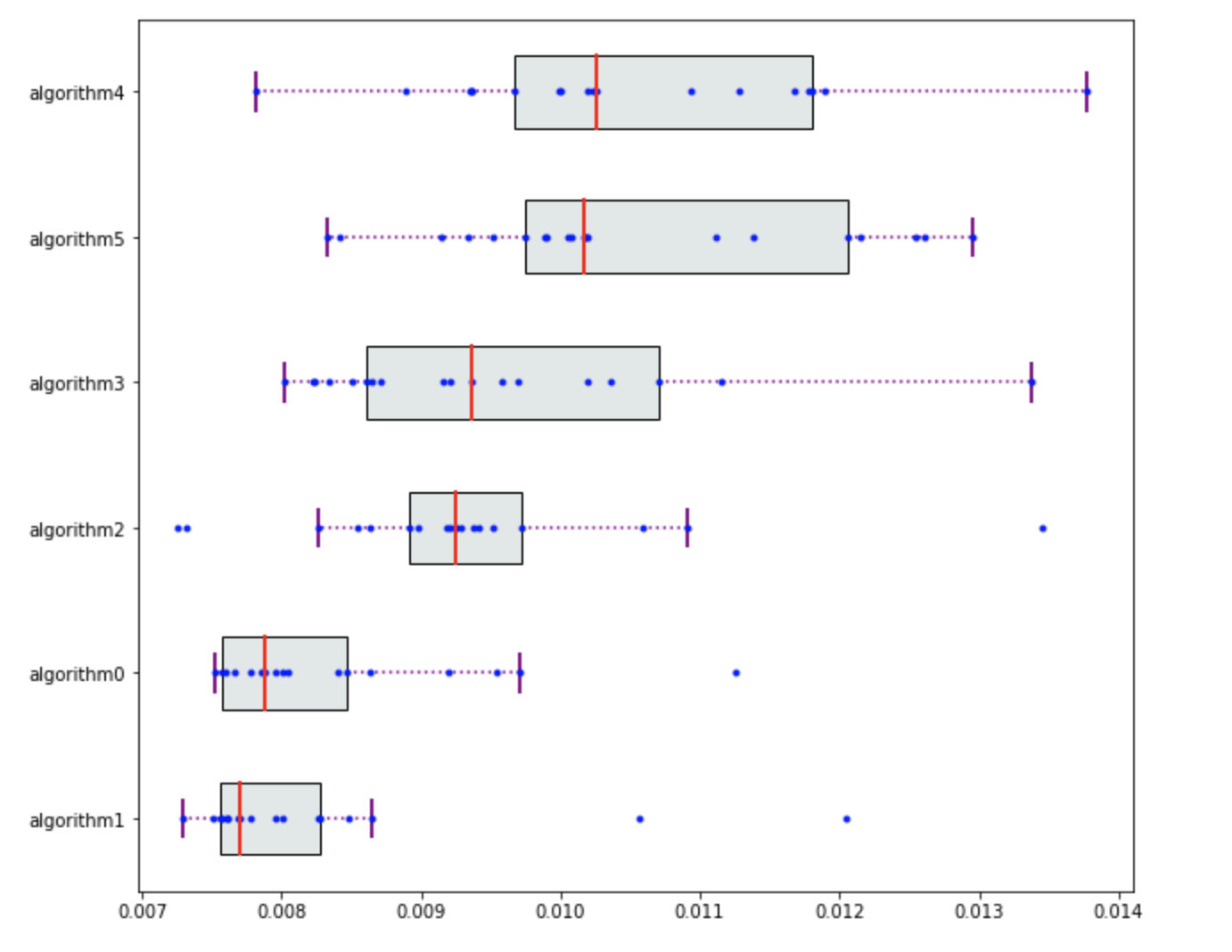}
		\caption{Instance A: Shared}
		\label{fig:front2b} 
	\end{subfigure}
	
	\begin{subfigure}[b]{0.5\textwidth}
		\includegraphics[width=1\linewidth]{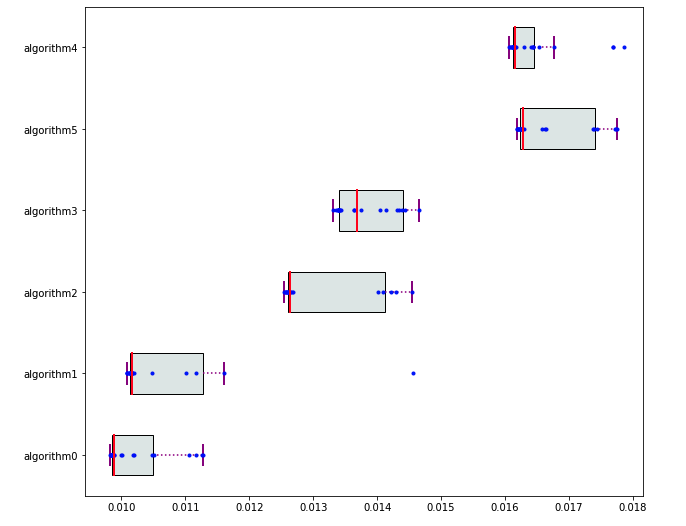}
		\caption{Instance A: Exclusive}
		\label{fig:turbo2b}
	\end{subfigure}
	
	\begin{subfigure}[b]{0.5\textwidth}
		\includegraphics[width=1\linewidth]{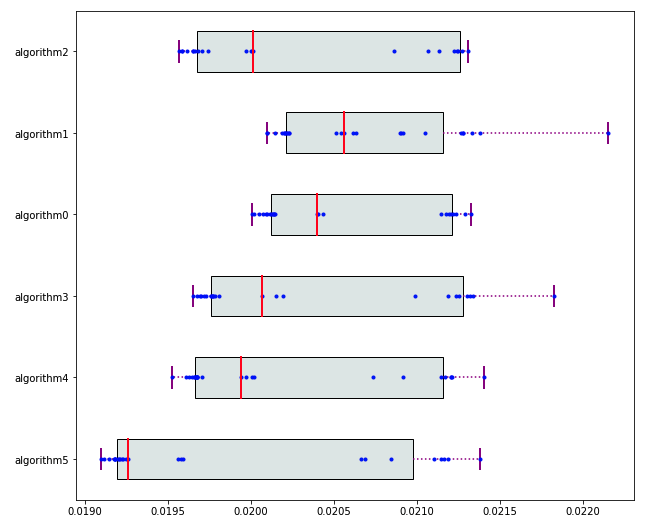}
		\caption{Instance B: Exclusive}
		\label{fig:turbo2c}
	\end{subfigure}
	
	\caption{Measurements for Instance A and Instance B. The algorithms are ordered according to the increasing order of their ranks from bottom to top. }
	\label{fig:2b}
\end{figure}

We consider the following instances of Expression~\ref{eqn:mc}: 
\begin{itemize}
	\item  Instance A: $(1000, 1000, 500, 1000, 1000 )$
	\item  Instance B: $(1000, 1000, 1000, 1000, 1000 )$  
\end{itemize}
For each instance, we first execute all the algorithms once. Then, the initial hypothesis ($\mathbf{h_0}$) is formed by ranking the algorithms in the increasing order of their single-run execution times. The Procedure~\ref{alg:measure} is applied with the parameters $M=3$, $eps=0.03$ and $max=30$. We define the quantile ranges same as those in Table~\ref{tab:q-ranks}. The experiments are run on two different settings. In the first setting, the experiments are run on a node, whose unused processing power and memory can be shared among other processes. In the second setting, the experiments are run on a node where exclusive access is granted. The execution times of the algorithms on the shared node is expected to have more fluctuations than the exclusive node. Before every iteration of the mean ranks computation in Procedure~\ref{alg:measure}, the $M$ measurements from every algorithm are shuffled to enable fair comparison. The results of the experiments are shown in Figure~\ref{fig:exp2}. The tables show the updated sequences, estimated ranks at $(q_{25}, q_{75})$ and the mean ranks. The shades of the cells indicate the relative FLOPs counts of the algorithms (see Equation~\ref{eq:rel-flops}); a darker shade indicates a higher relative FLOPs. The algorithms in white cells compute the least FLOPs.
\begin{itemize}
	\item \textbf{Instance A} (Figure~\ref{fig:exp2b}): The minimum FLOPs algorithms (\textbf{alg0} and \textbf{alg1}) obtain the best rank in both settings. In the shared setting, the mean rank of \textbf{alg0} is greater than \textbf{alg1}, which indicates that the underlying distribution of \textbf{alg0} is slightly shifted towards the right of \textbf{alg1} (see Figure~\ref{fig:front2b}); this indicates that  higher execution times are observed in some samples of \textbf{alg0} than in \textbf{alg1}. All the other algorithms obtain the same rank despite having different FLOP counts, which indicates significant overlap of distributions. However, the algorithms with the highest FLOP count (\textbf{alg4} and \textbf{alg5}) are slightly shifted to the right of \textbf{alg2} and \textbf{alg3}, indicating  relatively worse performance; this is quantified by the higher mean rank scores of \textbf{alg4} and \textbf{alg5} than \textbf{alg2} and \textbf{alg3}. On the other hand, in the exclusive setting, \textbf{alg4} and \textbf{alg5} obtain a higher rank than \textbf{alg2} and \textbf{alg3}, which indicates not only the rightward-shift but also a non-significant overlap of the underlying distributions (see Figure~\ref{fig:turbo2c}). The iterations in Procedure~\ref{alg:measure} stop after 21 and 24 measurements per algorithm in the shared and exclusive settings respectively.  
	
	\item \textbf{Instance B} (Figure~\ref{fig:exp2c}): All the algorithms compute comparable FLOPs and they all obtain the same rank in both the settings. In the shared mode, 15 measurements per algorithm were made, while the exclusive mode took 27 measurements per algorithm for the mean ranks to converge.  
\end{itemize}

\begin{figure}
	\centering
	
	\begin{subfigure}[b]{0.5\textwidth}
		\includegraphics[width=1\linewidth]{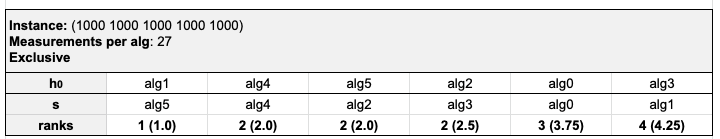}
		\caption{Instance B}
		\label{fig:anomaly2c}
	\end{subfigure}
	
	\begin{subfigure}[b]{0.5\textwidth}
		\includegraphics[width=1\linewidth]{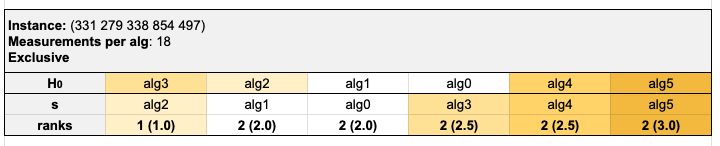}
		\caption{Anomaly}
		\label{fig:anomaly-1}
	\end{subfigure}
	
	\caption{Quantiles: $[(q_{5},q_{50}), (q_{15},q_{45}), (q_{20},q_{40}), (q_{25},q_{35}) ] $ The ranks at $(q_{45}, q_{15})$ and the mean ranks (in brackets) are shown.}
	\label{fig:anomaly}
\end{figure}

\textit{\textbf{Effect of Turbo boost:}} It can be noticed that in the exclusive setting, more measurements were made than in the shared setting. 
The scatter plots of algorithms in Figure~\ref{fig:turbo2b} and \ref{fig:turbo2c} show multi-modal distribution of measurements (especially bi-modal with two clusters of data points at the two ends of the distribution). This is because the processor operated at multiple frequency levels due to turbo boost settings, thereby resulting in significantly different execution times for the same algorithm. As the measurements of algorithms were sufficiently shuffled, the probability that a particular algorithm executes in just one frequency mode---thereby resulting in a biased comparison---is minimized. However, for the quantile ranges we considered (from Table~\ref{tab:q-ranks}), the algorithms \textbf{alg4}, \textbf{alg3}, \textbf{alg0}, \textbf{alg1}, \textbf{alg2} obtain the same mean rank scores (see exclusive mode in Figure~\ref{fig:exp2c}) even though the relative shifts among their distributions can be visually observed in Figure~\ref{fig:turbo2c}. In order to compare algorithms based on the measurements taken during the fast frequency modes of the processor (i.e., measurements towards the left end of the distribution), we modify the quantiles set in Procedure~\ref{alg:measure} and consider the following ranges: $[(q_{5},q_{50}), (q_{15},q_{45}), (q_{20},q_{40}), (q_{25},q_{35}) ]$ and recalculate the mean ranks. The results are shown in Figure~\ref{fig:anomaly2c}. Now, \textbf{alg5} obtains the best rank. The relative shifts among the algorithms based on the left-part of the distributions are now quantified by the mean ranks.  

\textit{\textbf{Test for FLOPs as a discriminant for the best algorithm:}} Consider the algorithms for instance B again. If the algorithms are to be executed in the compute node that operates at multiple frequency levels, then according to our methodology, at $(q_{25},q_{75})$, all the algorithms are considered equivalent, as they all obtain the best rank. The mean rank of \textbf{alg5} is better than the rest, but the methodology does not consider them statistically significant. Hence, one would not lose significantly in performance by randomly choosing one of the minimum FLOPs algorithm. However, if one is interested only in the performance at the high frequency modes of the processors, then \textbf{alg5} shows significantly better performance than the other algorithms. In this case, FLOPs fail to discriminate the algorithms and the instance will be considered as an anomaly. 

Recall the instance $(331, 279, 338, 854,  497)$ which was observed as an \textit{anomaly} in~\cite{Lopez2022:530} and discussed in Sec.~\ref{sec:int}. When the different frequency modes of the processors are not taken into account, then all the algorithms are equivalent as they all obtain rank 1. However, when focusing on the fast frequency modes, \textbf{alg2} (which is not the best algorithm based on FLOPs) shows significantly better performance than the algorithms with minimum FLOPs (see Figure~\ref{fig:anomaly-1}). Now, according to the methodology, this instance will be considered as an anomaly. 

Expression~\ref{eqn:mc} consisted of only 6 variants and we measured all of them in order to explain the working of the methodology. However, in practise, compilers such as Linnea generate 100s of variant algorithms for a given linear algebra expression. Then, one could filter the initial set of algorithms and create a subset consisting of only the potential algorithms before taking further measurements. In order to test if FLOPs are a valid discriminant for a given instance of an expression, the set of potential candidates could be all the algorithms with the least FLOP count and those algorithms whose relative times based on single-run execution times (calculated according to Equation~\ref{eq:rel-time}) are less than certain threshold (say, 1.5). Then, the Procedure~\ref{alg:measure} can be applied on the reduced set, consisting of only the potential algorithms. 

%
%

%


\section{Conclusion}
\label{sec:con}
In this work, we developed a methodology to rank a set of equivalent algorithms into performance classes. The input to the methodology is a set of algorithms ranked based on an initial hypothesis such as FLOP count or single measurement of execution time of each algorithm. We take further measurements of the algorithms in small steps incrementally, and update the ranks accordingly. The process of measurements stops as the updates to ranks converge. To this end, we developed a strategy to sort algorithms by comparing the quantiles of the execution time measurements; the ranks of the algorithms are merged if they have significant overlaps in the distribution of measurements. The rank estimates quantify the relative performances of the algorithms from one another. We showed that our methodology can be used to interpret and analyse performance even in compute nodes that operates at multiple frequency levels (e.g., machines that have turbo-boost enabled). We used our methodology to develop a test for FLOPs as a  discriminant for linear algebra algorithms. The Python implementation of the methodology is available online\footnote{https://github.com/as641651/AlgorithmRanking.}.

Recall our proposition (from Sec.~\ref{sec:int})  that high-level languages such as Julia, Matlab, TensorFlow, etc., choose algorithms that are sub-optimal in performance. The argument for the sub-optimality can be two folds: First, the languages do not fully apply the linear algebra knowledge to explore all possible alternate algorithms (this issue was not discussed in this paper). Second, they select a sub-optimal algorithm from a given set of alternatives; because these high-level languages make  algorithmic choices by minimizing FLOPs and it had been pointed out (e.g., in~\cite{Lopez2022:530}) that the algorithm with the lowest FLOP count is not always the fastest (such instances were referred to as \textit{anomalies}). In order to tackle the second argument, performance models that facilitate better algorithm selection strategy have to be developed; that is, those performance models should be able to perform better than what FLOPs can already do.  To this end, it is important to verify that, for a considered use case, there exists an abundance of anomalies that cannot be discriminated using FLOP counts. Our methodology can be used to detect the presence of anomalies. The anomalies can be used for further investigations to find the root-causes of performance differences. 


\section*{Acknowledgment}
Financial support from the Deutsche Forschungsgemeinschaft (German Research Foundation) through grants GSC 111 and IRTG 2379 is gratefully acknowledged. We thank Prof. Lars Karlsson from Ume\r{a} Universitet, Sweden for proof reading the paper and suggesting improvements. 

\bibliographystyle{IEEEtran}
\bibliography{sbac22}
\end{document}